\documentclass[aps,12pt,tightenlines,amsmath,amssymb,showpacs]{revtex4}
\usepackage{graphicx}% Include figure files
\usepackage{dcolumn}% Align table columns on decimal point
\usepackage{bm}% bold math

\begin{document}

\title{Comment on ``Irregularity in gamma ray source spectra as a signature of axion-like particles''}

\author{Giorgio Galanti}
\affiliation{Dipartimento di Fisica, Universit\`a dell'Insubria, Via Valleggio 11, I -- 22100 Como, Italy}
\email{gam.galanti@gmail.com}

\author{Marco Roncadelli}
\affiliation{INFN, Sezione di Pavia, Via A. Bassi 6, I -- 27100 Pavia, Italy}
\email{marco.roncadelli@pv.infn.it}

\begin{abstract}

D. Wouters and P. Brun in Phys. Rev. D {\bf 86}, 043005 (2012) claim that an observable effect in the spectra of distant very-high-energy blazars arises as a consequence of oscillations of photons into axion-like particles (ALPs) in the presence of turbulent extra-galactic magnetic fields. The main objective of this comment is to demonstrate that such a result is physically incorrect. We also show that a physically correct treatment of the same issue leads to a much less relevant conclusion, which makes the effect pointed out by WB likely unobservable with the present capabilities.

\end{abstract}


\maketitle

%%%%%%%%%%%%%%%%%%%%%%%%%%%%%%%%%%%%%%%%%%%%
%% MAINMATTER
%%%%%%%%%%%%%%%%%%%%%%%%%%%%%%%%%%%%%%%%%%%%

\section{INTRODUCTION}

Recently, Wouters and Brun (WB)~\cite{wb} proposed a new method to detect photon-ALP oscillations taking place in turbulent extra-galactic magnetic fields when very-high-energy observations of blazars are performed. 

Actually, the extra-galactic magnetic field ${\bf B}$ is supposed to have a domain-like structure, with its direction randomly changing from one domain to the next and strength either equal in all domains or with a Kolmogorov spectrum. For simplicity, we shall restrict our attention throughout this paper to the first option. Manifestly, in such a situation the propagation process of the photon/ALP beam from the source to us becomes a stochastic process. While it is obvious that the beam follows a single trajectory at once joining the source to us, the exact behavior of the beam cannot be predicted but only its mean properties can be evaluated, and this requires an average over a very large number of possible trajectories followed by the beam (realizations of the stochastic process in question). Among these properties, the simplest ones are the average photon survival probability~\cite{darma,dmpr,dgr} and its variance~\cite{mirmont}.

The main point made by WB is that in a pretty small range about the energy $E_{\rm thr}$ that marks the transition from the weak to the strong mixing regime the photon survival probability along every single trajectory that the beam can follow exhibits fluctuations, which show up in the observed energy spectrum and are claimed to be an {\it observable}  signature of the existence of photon-ALP oscillations. Incidentally, in the same energy range also the average photon survival probability oscillates~\cite{dmr}.

In our opinion, a flaw of the considered paper is that WB do not state explicitly their assumptions, neither they provide any information about the way they evaluate the photon survival probability along a single trajectory. Do they consider a polarized beam or an unpolarized one? As we will see below, this point is of crucial importance because it changes drastically the result, but nothing is said about that by WB. Quoting a famous statement of Georg Uhlenbeck ``first the assumptions, then the result''! 

We explicitly show that they consider an initially {\it polarized} beam, whereas a physically correct treatment demands the beam to be initially {\it unpolarized}. As a consequence, the result of WB changes {\it completely}.   

\section{SETTING THE STAGE}

Specifically, writing the photon-ALP Lagrangian as
\begin{equation}
\label{a1}
{\cal L}_{a \gamma} =  - \, \frac{1}{4} \, g \, F_{\mu\nu} \tilde{F}^{\mu\nu} a = g \, {\bf E} \cdot {\bf B} \, a~,
\end{equation}
the definition of $E_{\rm thr}$ is
\begin{equation}
\label{a2}
E_{\rm thr} \equiv \frac{|m_a^2 - \omega_{\rm pl}^2|}{2g \, B_T}~, 
\end{equation}
where $m_a$ is the ALP mass, $\omega_{\rm pl}$ is the plasma frequency and ${\bf B}_T$ is the component of the magnetic field transverse to the beam (WB write $B \, {\rm sin} \, \theta$ in place of $B_T$), which is supposed to be monochromatic of energy $E$. 

A very remarkable fact is that under the assumption $E \gg m_a$ the beam propagation equation in a generic magnetic domain $n$ takes a Schr\"odinger-like form in the variable $z$ along the beam~\cite{rs}, to wit
\begin{equation}
\label{a3}
\left( i \frac{d}{d z} + E + {\cal M} (\phi_n) \right) \, \psi_n (z) = 0
\end{equation}
with the wave function of the form
\begin{equation}
\label{a4}
\psi_n (z) \equiv \left(
\begin{array}{c}
A_{1,n} (z) \\
A_{2,n} (z) \\
a_n (z) \\
\end{array}
\right)~,
\end{equation}
where $A_{1,n} (z)$ and $A_{2,n} (z)$ denote the photon amplitudes with polarization (electric field) along the $x$- and $y$-axis, respectively, while $a_n (z)$ is the amplitude associated with the ALP in the $n$-th domain. Further, we let $\phi_n$ be the angle between ${\bf B}_T$ and the fixed ${\hat {\bf x}}$ direction -- equal for all domains -- in the $n$-th domain. In the general case in which the Extragalactic Background Light (EBL) is important a fraction of photons gets absorbed through the process $\gamma \gamma \to e^+ e^-$ and the mixing matrix ${\cal M} (\phi_n)$ reads
\begin{equation}
\label{a5}
{\cal M} (\phi_n) = \left(
\begin{array}{ccc}
{\Delta}_{11} + i \, {\Delta}_{\rm abs}& 0 & {\Delta}_B \,{\rm cos} \, \phi_n  \\
0 & {\Delta}_{22} + i \, {\Delta}_{\rm abs} & {\Delta}_B \,{\rm sin} \, \phi_n \\
{\Delta}_B \,{\rm cos} \, \phi_n & {\Delta}_B \, {\rm sin} \, \phi_n & \Delta_a \\
\end{array}
\right)~.
\end{equation}
The various delta terms are defined as follows: $\Delta_{11} = \Delta_{22} = - \, \omega^2_{\rm pl}/2 E$, 
$\Delta_a = - \, m^2_a/2 E$, $\Delta_B = g \, B_T/2$ and $\Delta_{\rm abs} = 1/2 \lambda$ with $\lambda$ being the photon means free path for $\gamma \gamma \to e^+ e^-$ scattering~\cite{cctp}. 

Hence, we see that inside every domain the considered beam is formally described as a three-level unstable 
non-relativistic quantum system. 

In the simplest case of a single domain with ${\bf B}$ homogeneous, 
$\omega_{\rm pl} = 0$, $\lambda = \infty$ and $\phi_n = \pi/2$ the $\gamma \to a$ conversion probability is
\begin{equation}
\label{a6}
P_{\gamma \to a} = \frac{\alpha {\Delta}^2_B}{{\Delta}^2_{\rm osc}} \, {\rm sin}^2 \left(\frac{{\Delta}_{\rm osc} \, z}{2} \right)~,
\end{equation}
having set ${\Delta}^2_{\rm osc} \equiv \Delta^2_a + 4 \Delta^2_B$. For a photon beam linearly polarized along 
${\bf B}_T$ we have $\alpha = 4$, for a linear polarization perpendicular to ${\bf B}_T$ we get $\alpha = 0$, whereas for an unpolarized beam it turns out that $\alpha = 2$. Moreover, Eq. (\ref{a6}) shows that for $E$ sufficiently larger than $E_{\rm thr}$ $P_{\gamma \to a}$ becomes maximal and energy-independent, which is indeed the strong mixing regime.

So, we see that the question whether the beam is polarized or not is of crucial importance because it {\it changes}  drastically the conclusion.

\section{PROBABILITIES FOR POLARIZED AND UNPOLARIZED BEAMS}

Our main criticism indeed concerns the beam polarization. We want to emphasize that the beam polarization is {\it unknown}. A reason is that it is not clear whether the emission mechanism is leptonic or hadronic, and for instance in the pure synchro-self-Compton model (without external electrons) the polarization of the emitted photons decreases both with the electron energy and the viewing angle, so that it is vanishingly small for the TeV BL Lacs~\cite{tav}. Another reason is that the polarization cannot be measured in the $\gamma$-ray band. So, in the lack of any information about the beam polarization the only sensible option is to suppose that the beam is initially {\it unpolarized}. 

Hence, according to quantum mechanics in the $n$-th domain the beam {\it must} be described by a polarization density matrix, namely
\begin{equation}
\label{a7}
\rho_n (z) = \left(\begin{array}{c}A_{n,1} (z) \\ A_{n,2} (z) \\ a_n (z)
\end{array}\right)
\otimes \Bigl(\begin{array}{c}A_{n,1} (z) \  A_{n,2} (z) \ a_n (z) \end{array}\Bigr)^{*}
\end{equation}
rather than by a wave function $\psi_n (z)$ like the one in Eq. (\ref{a4}). Moreover, the analogy with non-relativistic quantum mechanics entails that 
$\rho_n (z)$ obeys the Von Neumann-like equation
\begin{equation}
\label{a8}
i \frac{d \rho_n}{d z} = \rho_n \, {\cal M}^{\dagger} (\phi_n) - {\cal M} (\phi_n) \, \rho_n~.
\end{equation}
associated with Eq. (\ref{a3})~\cite{dgr}. Observe that even though the hamiltonian is not self-adjoint, we always have 
\begin{equation}
\label{a9}
\rho_n (z) = {\cal U}_n (z, 0) \, \rho_n (0) \, {\cal U}^{\dagger}_n (z, 0)~,
\end{equation}
where ${\cal U}_n (z, 0)$ is the transfer matrix, namely the solution of Eq. (\ref{a3}) subject to the initial condition ${\cal U}_n (0, 0) = 1$. Assuming that the number of domains is $N$, the transfer matrix describing the whole propagation process is
\begin{equation}
\label{a10}
{\cal U} (z, 0) = \prod^{N}_{n = 1} \, {\cal U}_n (z_n, z_{n - 1})
\end{equation}
with $z_0 = 0$ and $z_N = z$, and the photon survival probability along a {\it single} realization of the {\it unpolarized} beam corresponding to $\phi_1, \phi_2, . . . , \phi_N$ is given by~\cite{dgr}
\begin{equation}
\label{a11}
P_{\gamma \to \gamma}^{\rm unpolarized} (z, 0; \phi_1, \phi_2, . . . , \phi_N) = \sum_{i = 1,2} {\rm Tr} \Bigl( \rho_i \, {\cal U} (z, 0) \, \rho_{\rm unpol} \, {\cal U}^{\dagger} (z, 0) \Bigr)~,
\end{equation}
while the analogous probability in the case of a {\it polarized} beam along the $x$-axis reads~\cite{remarkQ}
\begin{equation}
\label{a11q}
P_{\gamma \to \gamma}^{\rm polarized} (z, 0; \phi_1, \phi_2, . . . , \phi_N) = \sum_{i = 1,2} {\rm Tr} \Bigl( \rho_i \, {\cal U} (z, 0) \, \rho_1 \, {\cal U}^{\dagger} (z, 0) \Bigr)~,
\end{equation}
where 
\begin{equation}
\label{a12}
{\rho}_1 = \left(
\begin{array}{ccc}
1 & 0 & 0 \\
0 & 0 & 0 \\     
0 & 0 & 0 \\
\end{array}
\right)~, \ \ \ 
{\rho}_2 = \left(
\begin{array}{ccc}
0 & 0 & 0 \\
0 & 1 & 0 \\
0 & 0 & 0 \\
\end{array}
\right)~, \ \ \ 
{\rho}_{\rm unpol} = \frac{1}{2}\left(
\begin{array}{ccc}
1 & 0 & 0 \\
0 & 1 & 0 \\
0 & 0 & 0 \\
\end{array}
\right)~.
\end{equation}

\section{A PARTICULAR CASE}

Let us consider first the case in which EBL absorption is absent, so that the hamiltonian is self-adjoint and the transfer matrix must be unitary. Since we cannot know the specific trajectory followed by the beam during its propagation, this has to be true for {\it any} trajectory. Now, by inserting Eqs. (\ref{a12}) into Eq. (\ref{a11}) and working out the resulting expression we find
\begin{equation}
\label{a13}
P_{\gamma \to \gamma} (z, 0; \phi_1, \phi_2, . . . , \phi_N) = \frac{1}{2} \Bigl(|u_{11}|^2 + |u_{12}|^2 + |u_{21}|^2 + |u_{22}|^2 \Bigr)~.
\end{equation}
But owing to the unitarity of ${\cal U}$, the condition ${\cal U} \, {\cal U}^{\dagger} = 1$ implies
\begin{equation}
\label{a14}
|u_{11}|^2 + |u_{12}|^2 + |u_{13}|^2 = 1~, \ \ |u_{21}|^2 + |u_{22}|^2 + |u_{23}|^2 = 1~, \ \ |u_{31}|^2 + |u_{32}|^2 + |u_{33}|^2 = 1~,
\end{equation}
whereas the condition ${\cal U}^{\dagger} \, {\cal U} = 1$ entails
\begin{equation}
\label{a15}
|u_{11}|^2 + |u_{21}|^2 + |u_{31}|^2 = 1~, \ \ |u_{12}|^2 + |u_{22}|^2 + |u_{32}|^2 = 1~, \ \ |u_{13}|^2 + |u_{23}|^2 + |u_{33}|^2 = 1~,
\end{equation}
which upon insertion into Eq. (\ref{a13}) yield
\begin{equation}
\label{a17}
P_{\gamma \to \gamma} (z, 0; \phi_1, \phi_2, . . . , \phi_N) = \frac{1}{2} + \frac{1}{2} \, |u_{33}|^2 \geq \frac{1}{2}~.
\end{equation}
This conclusion is in blatant contradiction with the result of WB reported in the upper panel of their Fig. 2, and so we infer that WB consider an initially {\it polarized} beam.

\section{PHOTON SURVIVAL PROBABILITY}

From now on we address the case in which the EBL absorption is present. 

\begin{figure}[h]
\centering
\includegraphics[width=1.20\textwidth]{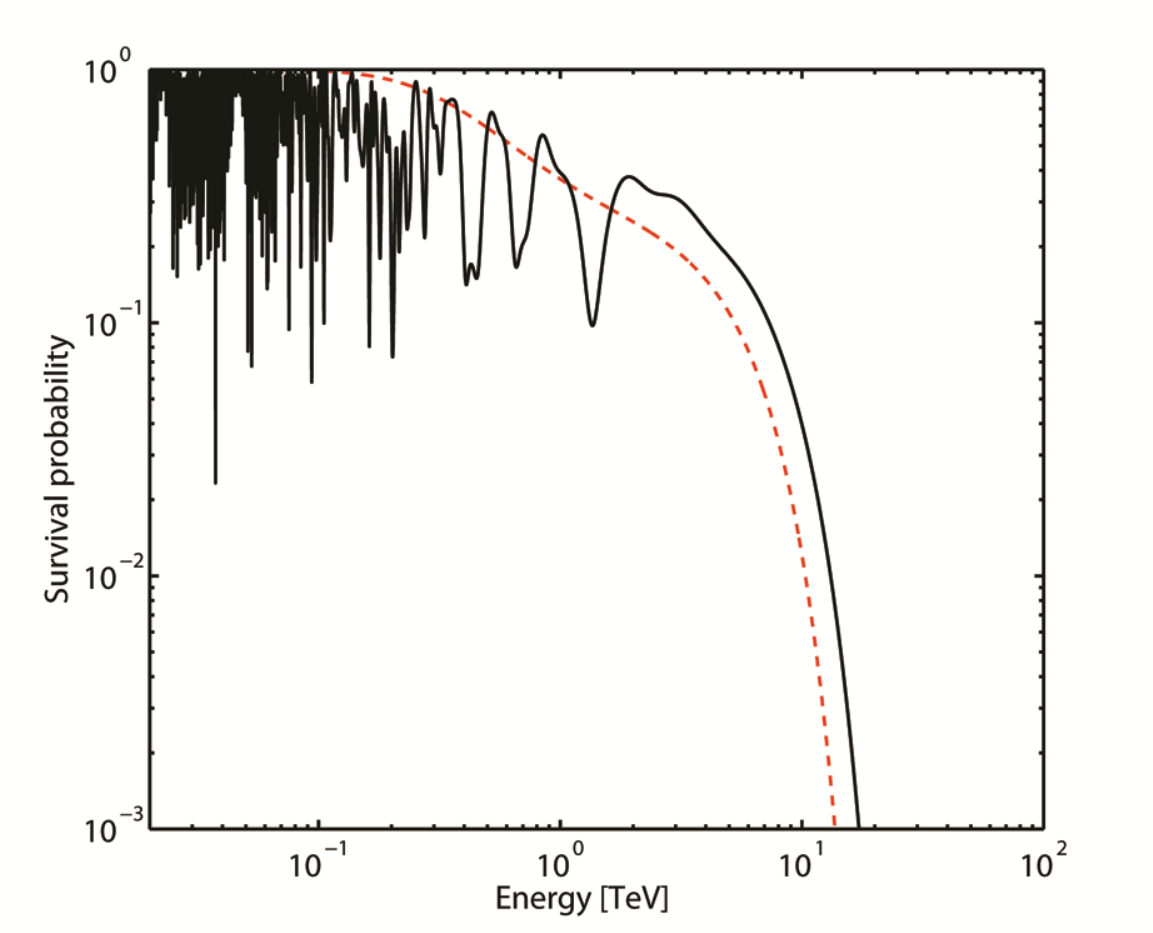}
\caption{\label{fig:pol} The solid black line represents the photon survival probability along a single randomly chosen trajectory followed by the beam in the case of an initially {\it polarized} beam in the presence of ALPs for a source at 
$z_s = 0.1$, using $B = 1 \, {\rm nG}$, the size of a magnetic domain equal to $1 \, {\rm Mpc}$, $g = 8 \cdot 10^{- 11} \, {\rm GeV}^{- 1}$ and $m_a = 2 \, {\rm neV}$. The dashed red line represents the same quantity with ALPs discarded (conventional physics).}
\end{figure}

As a benchmark for comparison, we start by dealing with the photon survival probability along a single randomly 
chosen trajectory of the considered stochastic process in the case of an initially {\it polarized} beam. For the sake of comparison with WB, we take the same values of the parameters adopted by them, namely a source at redshift $z_s = 0.1$ (not to be confused with the coordinate along the beam), the magnetic field strength $B = 1 \, {\rm nG}$, the size of a magnetic domain equal to $1 \, {\rm Mpc}$, the photo-ALP coupling $g = 8 \cdot 10^{- 11} \, {\rm GeV}^{- 1}$ and the ALP mass $m_a = 2 \, {\rm neV}$. Using Eq. (\ref{a11q}), we find the result plotted in Fig.~\ref{fig:pol}. Manifestly Fig.~\ref{fig:pol} is qualitatively identical to the lower panel of Fig. 2 of WB. This circumstance confirms that WB indeed consider an initially {\it polarized} beam.

\begin{figure}[h]
\centering
\includegraphics[width=1.20\textwidth]{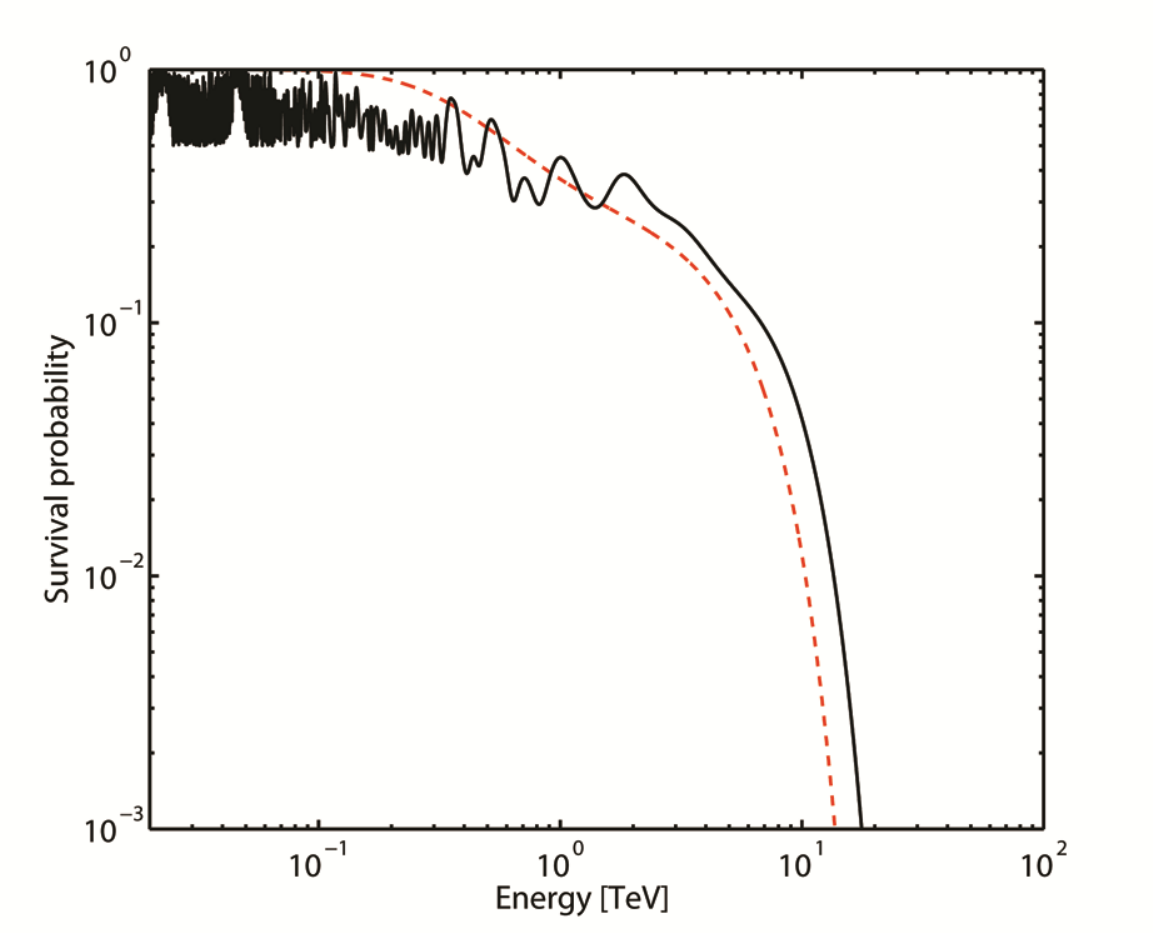}
\caption{\label{fig:unpol} The solid black line represents the photon survival probability along a single randomly chosen realization of the beam propagation in the case of an initially {\it unpolarized} beam in the presence of ALPs for a source at $z_s = 0.1$, using $B = 1 \, {\rm nG}$, the size of a magnetic domain equal to $1 \, {\rm Mpc}$, $g = 8 \cdot 10^{- 11} \, {\rm GeV}^{- 1}$ and $m_a = 2 \, {\rm neV}$. The dashed red line represents the same quantity with ALPs discarded (conventional physics).}
\end{figure}

Let us next address the analogous probability -- for the same values of the parameters -- in the case of an initially {\it unpolarized} beam, which is the physically correct case. Employing now Eq. (\ref{a11}), the corresponding result is exhibited in Fig.~\ref{fig:unpol}.

Evidently the size of the fluctuations is {\it drastically reduced} in the unpolarized case with respect to the polarized one.

\begin{figure}[h]
\centering
\includegraphics[width=1.20\textwidth]{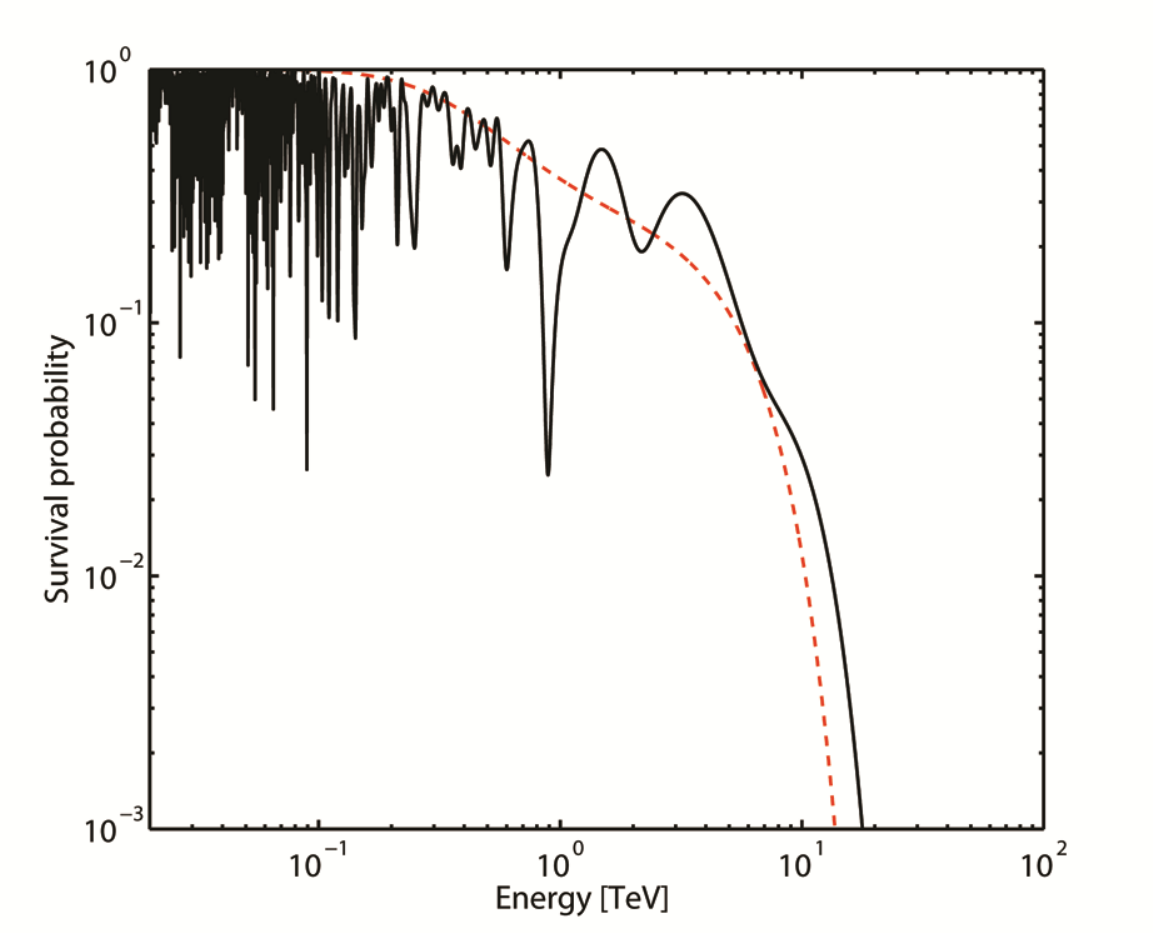}
\caption{\label{fig:pol2} The solid black line represents the photon survival probability along another single randomly chosen trajectory followed by the beam in the case of an initially {\it polarized} beam in the presence of ALPs for a source at $z_s = 0.1$, using $B = 1 \, {\rm nG}$, the size of a magnetic domain equal to $1 \, {\rm Mpc}$, $g = 8 \cdot 10^{- 11} \, {\rm GeV}^{- 1}$ and $m_a = 2 \, {\rm neV}$. The dashed red line represents the same quantity with ALPs discarded (conventional physics).}
\end{figure}

\begin{figure}[h]
\centering
\includegraphics[width=1.20\textwidth]{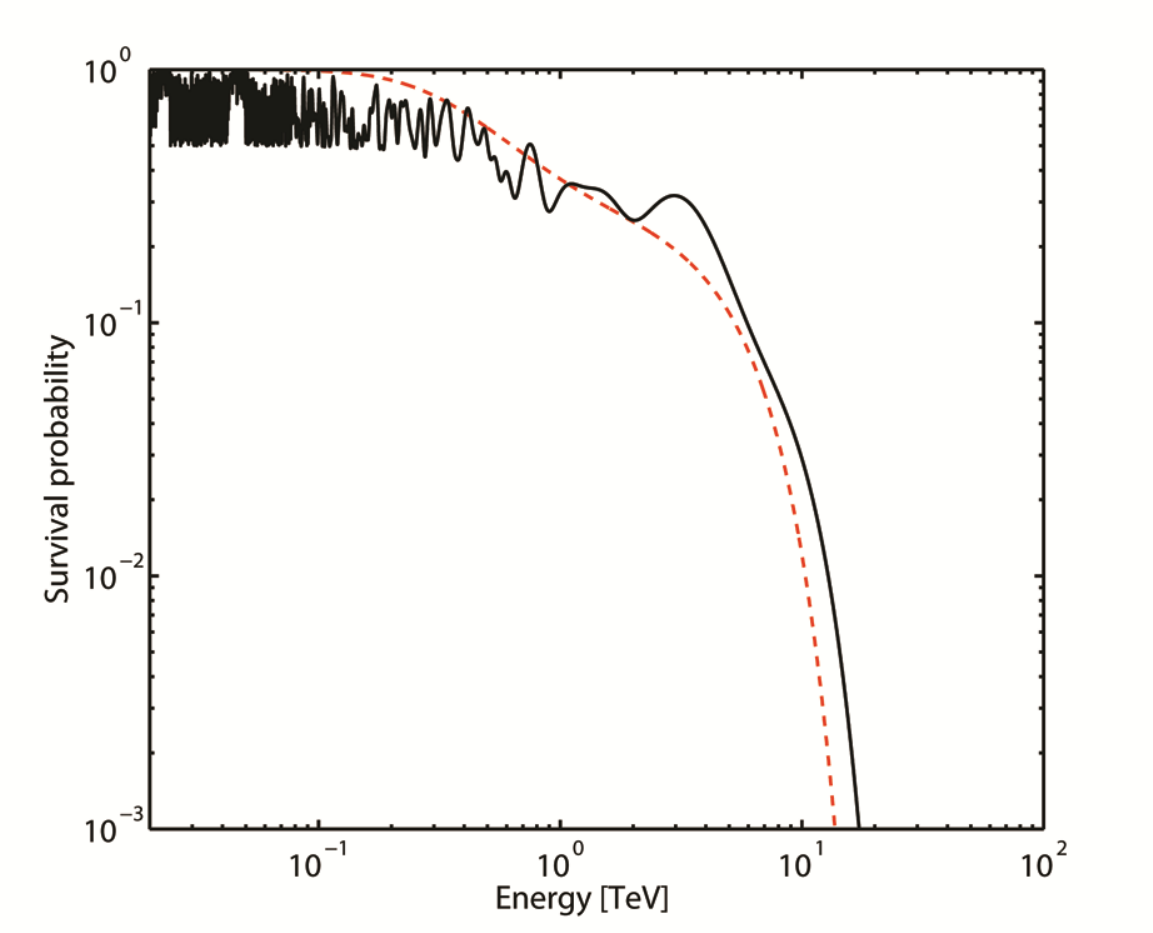}
\caption{\label{fig:unpol2} The solid black line represents the photon survival probability along another single randomly chosen realization of the beam propagation in the case of an initially {\it unpolarized} beam in the presence of ALPs for a source at $z_s = 0.1$, using $B = 1 \, {\rm nG}$, the size of a magnetic domain equal to $1 \, {\rm Mpc}$, $g = 8 \cdot 10^{- 11} \, {\rm GeV}^{- 1}$ and $m_a = 2 \, {\rm neV}$. The dashed red line represents the same quantity with ALPs discarded (conventional physics).}
\end{figure}

Just to show that such a conclusion is general -- and not a particular feature of the selected trajectory -- we take another randomly chosen trajectory. Repeating the above calculations in this case, the result for an initially polarized beam is plotted in Fig. \ref{fig:pol2}, while the one for an initially unpolarized beam is reported in Fig.~\ref{fig:unpol2}. Manifestly, Figs.~\ref{fig:pol} and \ref{fig:pol2} are qualitatively identical, and the same is true for Figs.~\ref{fig:unpol} and \ref{fig:unpol2}.

\section{OBSERVED FLUX}

As a further step, we follow as closely as possible the same lines of Sect. III of WB. Explicitly, as a first step we generate photons by a Monte Carlo method according to a log-parabola probability distribution -- shape of the initial spectrum -- with an integrated flux in the TeV band at the Crab level. We simulate an observation of 50 h with an effective area of $10^5 \, {\rm m}^2$, which amounts to about 100000 photons. We suppose that 10 observations of 5 h each are performed, so that every one collects about 10000 photons. Assuming that the observations are performed in the energy band $5 \cdot 10^2 \, {\rm GeV} - 7 \, {\rm TeV}$, we divide this range into 33 energy bins. At this point, we bin the 10  observations, computing both the mean and the variance pertaining to the 10 observations for each of the 33 energy bins. Next, we perform a log-log best fit of the binned points and we evaluate the fit residuals. Finally, we compute the variance of the fit residuals. All this is obtained by averaging over 5000 realizations, as in the case of WB. 

\begin{figure}[h]
\centering
\includegraphics[width=1.20\textwidth]{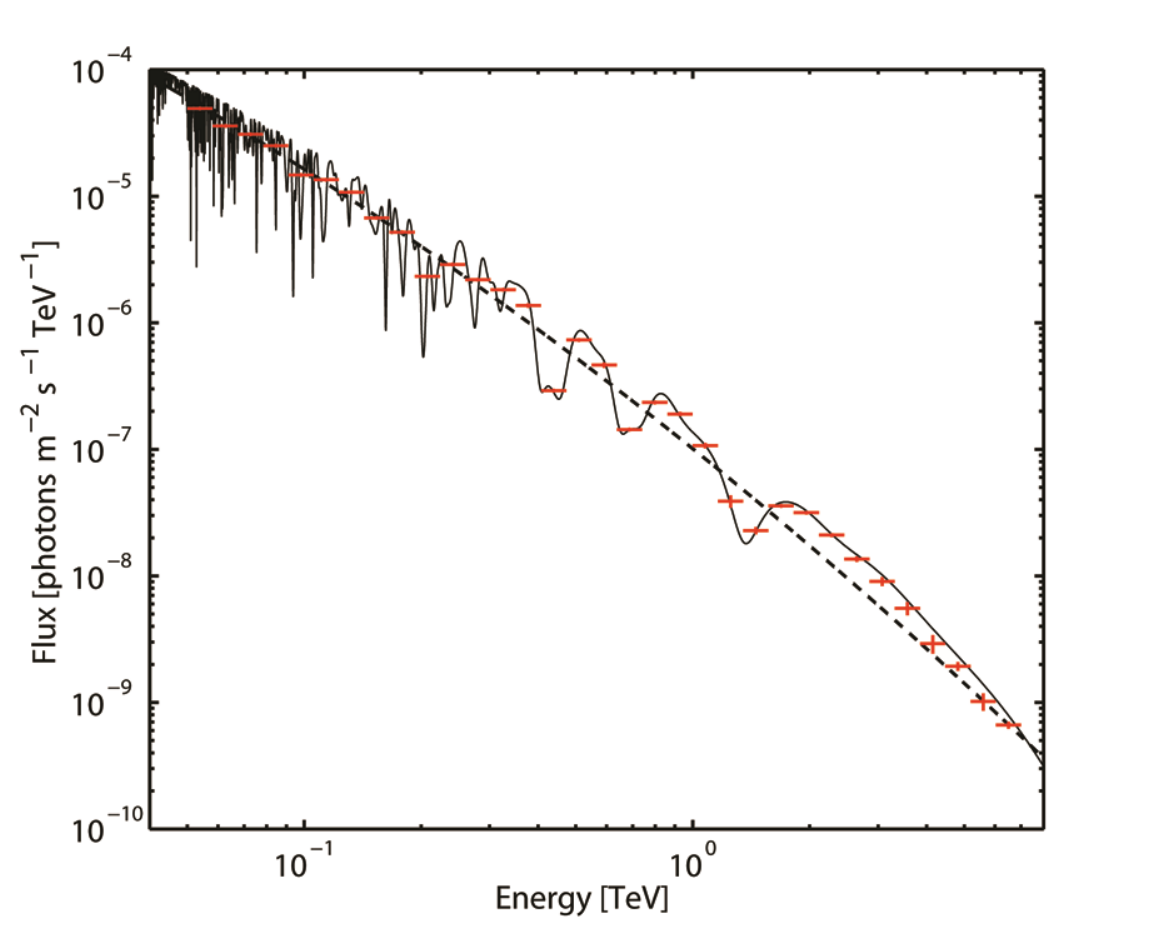}
\caption{\label{fig:fluxpol} This figure corresponds to the case of a polarized beam. The solid black line represents the unbinned spectrum and the red lines the binned spectrum in the case of photon-ALP oscillations. The dashed black line corresponds to the best fit to the bins (regardless of the underlying physics).}
\end{figure}

\begin{figure}[h]
\centering
\includegraphics[width=1.20\textwidth]{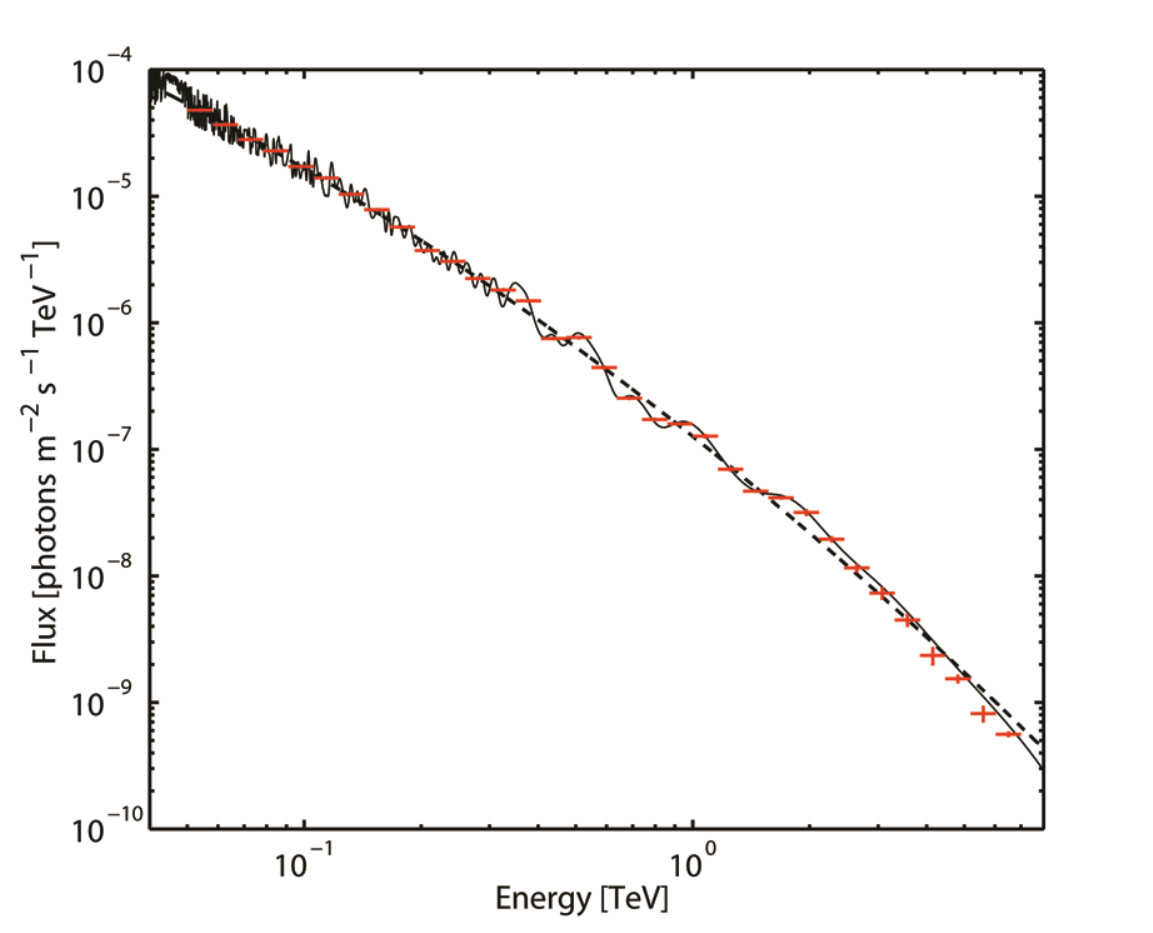}
\caption{\label{fig:fluxunpol} This figure corresponds to the case of an unpolarized beam. The solid black line represents the unbinned spectrum and the red lines the binned spectrum in the case of photon-ALP oscillations. The dashed black line corresponds to the best fit to the bins (regardless of the underlying physics).}
\end{figure}

We proceed in parallel with the discussion in Sect. V, which amounts to implement such a strategy first the case of an initially {\it polarized} beam and next the case of an initially {\it unpolarized} one. We show in Fig.~\ref{fig:fluxpol} the unbinned and binned spectra in the case of a {\it polarized} beam when EBL absorption and photon-ALP oscillations are considered. The model parameters are the same as before. Fig.~\ref{fig:fluxunpol} is merely the counterpart of Fig.~\ref{fig:fluxpol} in the case of an {\it unpolarized} beam. In either case, the solid black line represents the unbinned spectrum and the red lines the binned spectrum in the situation of photon-ALP oscillations. The dashed black line corresponds to the best fit to the bins (regardless of the underlying physics).

\begin{figure}[h]
\centering
\includegraphics[width=1.20\textwidth]{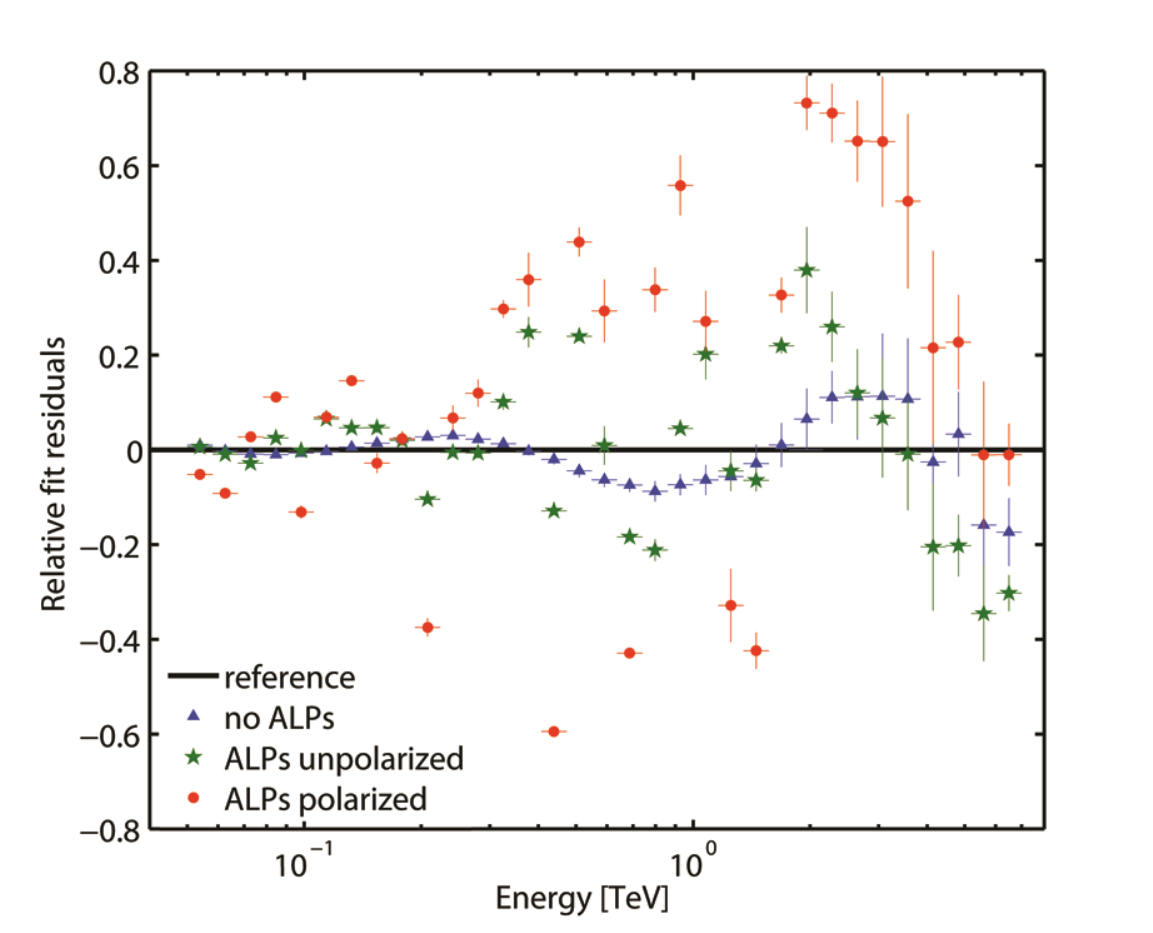}
\caption{\label{fig:resULT} Distribution of the residuals in three cases. Polarized beam with ALP effects: red blobs. Unpolarized beam with ALP effects: green stars. Conventional physics: blue triangles.}
\end{figure}

As before, the difference between Fig.~\ref{fig:fluxpol} and Fig.~\ref{fig:fluxunpol} is great: while in the polarized case the amplitude of the oscillations is large, in the unpolarized one their size gets drastically reduced. The actual physical difference between the two cases is confirmed by the distribution of the residuals -- displayed in Fig.~\ref{fig:resULT} -- where red blobs and green stars represent the cases of a polarized and unpolarized beam, respectively in the presence of photon-ALP oscillations. For comparison, the blue triangles correspond to the situation of conventional physics.

Finally, we report in Table~\ref{tab:var} the predicted values of the variance of the fit residuals for the above choice of the model parameters.

\begin{table}
\begin{tabular}{rr}
\hline
 {Model   \ \ \ \ \ \ \ \ \ \ \ \ \ \ \ \    }{}   
&  {}{Variance of the fit residuals} \\
\hline
   No ALPs  \ \ \ \ \ \ \ \ \ \ \ \ \ \ \        & $0.03 \pm 0.01 $  \ \ \ \ \ \ \ \ \ \ \ \  \\
   ALPs unpolarized  \ \ \ \ \ \ \ \ \ \         & $0.09 \pm 0.03$  \ \ \ \ \ \ \ \ \ \ \ \  \\
   ALPs polarized   \ \ \ \ \ \ \ \ \ \ \       & $0.21 \pm 0.06$  \ \ \ \ \ \ \ \ \ \ \ \  \\
\hline
\end{tabular}
\caption{Values of the variance to the fit residuals for the various cases considered in the text.}
\label{tab:var}
\end{table}

\section{CONCLUSIONS}

We have critically analyzed the claim put forward by WB~\cite{wb} that an observable effect in the spectra of distant very-high-energy blazars arises as a consequence of oscillations of photons into axion-like particles (ALPs) in the presence of turbulent extra-galactic magnetic fields. In practice, we have redone the same analysis of WB in order to understand whether their result concerning potentially observable fluctuations in the spectra of blazars in the  presence of photon-ALP oscillations are derived for a polarized or unpolarized photon/ALP beam. We have reproduced all their results in the case of an initially {\it polarized} beam, which however looks physically irrelevant to current observations. But we have shown that for the physically relevant case of an initially {\it unpolarized} beam the claimed effect is drastically reduced, indeed to such an extent to become likely unobservable with the present capabilities. In this respect, two remarks are in order. We have taken an energy resolution of $15 \, \%$ in order to conform ourselves with the choice of WB, but we believe that while this figure is realistic for the Cherenkov Telescope Array (CTA) it is too optimistic for the present Imaging Atmospheric Cherenkov Telescopes (IACTs), for which a value of $20 \, \%$ would be more realistic:  this would lead to a larger smearing of the fluctuations in the energy spectrum. An additional smearing arises from the systematic errors, which have not been taken into account again in order to conform our analysis with that of WB.

\section*{Acknowledgments}

We thank Alessandro De Angelis, Massimo Dotti, Emanuele Ripamonti and Fabrizio Tavecchio for discussions. M. R. acknowledges the INFN grant FA51.

\newpage

\newpage

\newpage

\newpage

\newpage

\newpage

\newpage

\end{document}